\documentclass[a4paper,onecolumn,showpacs,notitlepage,floatfix,nofootinbib,superscriptaddress,prd]{revtex4-1}
\usepackage{graphicx}
\usepackage{amsfonts}
\usepackage{amsmath}
\usepackage{hyperref}
\usepackage{float}
\usepackage{amssymb,amsbsy}
\usepackage{epstopdf}
\usepackage{color}
\usepackage{oldgerm}

\begin{document}

\title{Resistive dissipative magnetohydrodynamics from the Boltzmann-Vlasov equation}

\author{Gabriel S.\ Denicol}
\affiliation{Instituto de F\'isica, Universidade Federal Fluminense, UFF,
Niter\'oi, 24210-346, RJ, Brazil}

\author{Etele Moln\'ar}
\affiliation{Institut f\"ur Theoretische Physik, 
Johann Wolfgang Goethe--Universit\"at,
Max-von-Laue-Str.\ 1, D--60438 Frankfurt am Main, Germany} 
\affiliation{Institute of Physics and Technology, University of Bergen,
Allegaten 55, 5007 Bergen, Norway}

\author{Harri Niemi}
\affiliation{Institut f\"ur Theoretische Physik, 
Johann Wolfgang Goethe--Universit\"at,
Max-von-Laue-Str.\ 1, D--60438 Frankfurt am Main, Germany} 
\affiliation{Department of Physics, University of Jyv\"askyl\"a, P.O.\ Box
35, FI-40014 University of Jyv\"askyl\"a, Finland} 
\affiliation{Helsinki
Institute of Physics, P.O.\ Box 64, FI-00014 University of Helsinki, Finland}

\author{Dirk H.\ Rischke}
\affiliation{Institut f\"ur Theoretische Physik, 
Johann Wolfgang Goethe--Universit\"at,
Max-von-Laue-Str.\ 1, D--60438 Frankfurt am Main, Germany} 
\affiliation{Department of
Modern Physics, University of Science and Technology of China, Hefei, Anhui 230026, China}


\begin{abstract}
We derive the equations of motion of relativistic, resistive, second-order
dissipative magnetohydrodynamics from the Boltzmann-Vlasov equation using
the method of moments. We thus extend our previous work [Phys.\ Rev.\ D \textbf{98}, 076009 (2018)], 
where we only considered the non-resistive limit, to the case of finite electric conductivity. 
This requires keeping terms proportional to the electric field $E^\mu$ in the equations of motions
and leads to new transport coefficients due to the coupling of the electric field to
dissipative quantities. We also show that the Navier-Stokes limit of the
charge-diffusion current corresponds to Ohm's law, while the coefficients
of electrical conductivity and charge diffusion are related by a type of Wiedemann-Franz law.
\end{abstract}

\maketitle

\section{Introduction}

Second-order theories of relativistic dissipative fluid dynamics play an essential role in understanding the dynamics of 
ultrarelativistic heavy-ion collisions~\cite{Heinz:2013th}. Moreover, strong electromagnetic fields are created in 
non-central heavy-ion collisions~\cite{Skokov:2009qp,Deng:2012pc,Tuchin:2013apa,Bzdak:2011yy}, which give
rise to novel and interesting phenomena in strongly interacting matter, like the Chiral Magnetic Effect 
[for a review, see Ref.\ \cite{Huang:2015oca} and refs.\ therein].
In order to describe the evolution of the system, second-order relativistic dissipative
fluid dynamics~\cite{Israel:1979wp, Denicol:2012cn} needs to be extended to a self-consistent magnetohydrodynamic 
framework~\cite{degroot,Rezzolla_book:2013}.

In Ref.~\cite{Denicol:2018rbw} the equations of motion of relativistic, non-resistive, 
second-order dissipative magnetohydrodynamics were derived from the Boltzmann-Vlasov equation. 
In a non-resistive, i.e., ideally conducting, fluid the electric field is not an independent degree of freedom but
is related to the magnetic field by $\mathbf{E}=-\mathbf{v}\times \mathbf{B}$ and therefore can be eliminated 
from the equations of motion. While this is a common approximation in magnetohydrodynamics, 
it cannot be realized in a fully consistent manner in a system whose microscopic dynamics is described by the 
Boltzmann equation. The reason is that the electric conductivity $\sigma_E$ 
is a fluid-dynamical transport coefficient and thus, like
all other transport coefficients, proportional to the mean free path of the particles. Taking the limit 
$\sigma_E \rightarrow \infty$ while keeping the values of all other transport coefficients finite is inconsistent.

In this paper we will dispense with the assumption of infinite conductivity, and derive the equations of motion of resistive, 
second-order dissipative magnetohydrodynamics. As in our previous work~\cite{Denicol:2018rbw} we assume a 
single-component system of spin-zero particles with electric charge $\textswab{q}$ undergoing binary elastic collisions. 
The fluid-dynamical equations of motion are derived by using the 14-moment approximation in the framework developed 
in Refs.\ \cite{Denicol:2010xn,Denicol:2012cn,Denicol:2012es}. 
The electric field is now included explicitly, and the resistive magnetohydrodynamic equations of motion contain new terms 
with new transport coefficients due to the coupling of charged particles to the electric field.

The electric conductivity $\sigma _{E}$ is defined through
Ohm's law of magnetohydrodynamics, $\textswab{J}_{ind}^{\mu } = \sigma_{E}E^{\mu }$,
where $\textswab{J}_{ind}^{\mu}$ is the charge current induced by the electric field $E^{\mu }$.
We will show that the electric conductivity is related to the thermal conductivity $\kappa$, giving
rise to a type of Wiedemann-Franz law, 
$\sigma_{E}\equiv \textswab{q}^2 \kappa/T$, where $T$ is the temperature of matter. 

The paper is organized as follows. In Sec.\ \ref{sec:MHD} we recall 
the equations of motion of magnetohydrodynamics. In Sec.\ \ref{sec:diss_eom_moments} we derive the 
infinite set of equations of motion for the irreducible moments up to tensor-rank two of the deviation of the
single-particle distribution function from local equilibrium.
Section \ref{sec:diss_eom_2ndorder} is devoted to truncating this infinite system applying the 14-moment approximation, 
to obtain the equations for resistive, second-order dissipative
magnetohydrodynamics. The Navier-Stokes limit of these equations is discussed 
in Sec.\ \ref{NS_theory}. The last section contains a summary of this work.

We adopt natural Heaviside-Lorentz units $\hbar =c=k_{B}=\epsilon _{0}=\mu_{0}=1$, 
and the Minkowski space-time metric $g^{\mu \nu }=\text{diag}(1,-1,-1,-1)$. 
The fluid four-velocity is $u^{\mu } =\gamma \left( 1,\mathbf{v}\right) ^{T}$, 
with $\gamma =(1-\mathbf{v}^{2})^{-1/2}$ and normalization $u^{\mu }u_{\mu }\equiv 1$, 
while in the local rest (LR) frame of the fluid, $u_{LR}^{\mu }=\left( 1,\mathbf{0}\right) ^{T}$. 
The rank-two projection operator onto the three-space
orthogonal to $u^{\mu }$ is defined as $\Delta ^{\mu \nu }\equiv g^{\mu \nu}-u^{\mu }u^{\nu }$. 
For any four-vector, $A^{\mu }$, we define its
projection onto the three-dimensional subspace orthogonal to $u^{\mu }$ as 
$A^{\left\langle \mu \right\rangle }\equiv \Delta _{\nu }^{\mu }A^{\nu }$. A
straightforward generalization is the symmetric and
traceless projection tensor of rank-$2\ell $, denoted by 
$\Delta _{\nu _{1}\cdots \nu _{\ell }}^{\mu _{1}\cdots \mu _{\ell }}$, such
that the irreducible projections are $A^{\left\langle \mu _{1}\cdots \mu_{\ell }\right\rangle }
\equiv \Delta _{\nu _{1}\cdots \nu _{\ell }}^{\mu_{1}\cdots \mu _{\ell }}A^{\nu _{1}\cdots \nu _{\ell }}$ \cite{deGroot}.
As an example, the rank-four symmetric and traceless projection operator
is defined as $\Delta_{\alpha \beta }^{\mu \nu }\equiv \frac{1}{2}\left(
\Delta _{\alpha }^{\mu }\Delta _{\beta }^{\nu }+\Delta _{\beta }^{\mu}\Delta _{\alpha }^{\nu }\right) 
-\frac{1}{3}\Delta ^{\mu \nu }\Delta_{\alpha \beta }$.

The four-momentum $k^{\mu }$ of particles is normalized to their rest mass
squared, $k^{\mu }k_{\mu }=m_{0}^{2}$. The energy of a particle in the
LR frame of the fluid is defined as $E_{\mathbf{k}}\equiv k^{\mu }u_{\mu }$ and
coincides with the on-shell energy $k^{0}=\sqrt{\mathbf{k}^{2}+m_{0}^{2}} $.
The three-momentum of particles, $\mathbf{k}$, is defined through the
orthogonal projection of the four-momenta, $k^{\left\langle \mu
\right\rangle }\equiv \Delta _{\nu }^{\mu }k^{\nu }$, in the LR frame. The
comoving derivative of a quantity $A$ is denoted by an overdot, i.e., 
$\dot{A}\equiv u^{\mu }\partial _{\mu }A$, while the three-space gradient is 
$\nabla_{\nu }A\equiv \Delta_{\nu }^{\alpha }\partial _{\alpha}A$, 
hence in the LR frame they reduce to the usual time and spatial
derivatives $\partial _{t}A$ and ${\boldmath\textrm{$\nabla$}} A$. Furthermore, we use the decomposition
$\partial _{\mu }u_{\nu } = u_{\mu }\dot{u}_{\nu }+\frac{1}{3}\theta \Delta _{\mu \nu }+\sigma _{\mu\nu }
+\omega _{\mu \nu }$, 
where we define the expansion scalar, $\theta \equiv \nabla _{\mu }u^{\mu }$, the shear 
tensor $\sigma ^{\mu \nu }\equiv \nabla^{\left\langle \mu \right. }u^{\left. \nu \right\rangle }
=\frac{1}{2}(\nabla^{\mu }u^{\nu }+\nabla ^{\nu }u^{\mu })-\frac{1}{3}\theta \Delta ^{\mu \nu
}$, and the vorticity $\omega ^{\mu \nu }\equiv \frac{1}{2}(\nabla ^{\mu}u^{\nu }-\nabla ^{\nu }u^{\mu })$.

\section{Equations of motion of magnetohydrodynamics}
\label{sec:MHD}

The equations of motion of magnetohydrodynamics are [see Eqs.\ (24) and (25) of Ref.\ \cite{Denicol:2018rbw}]
\begin{eqnarray}
\partial _{\mu }\textswab{J}_{f}^{\mu } &=&0\;,  \label{J_f_mucons} \\
\partial _{\nu }T^{\mu \nu } &=&-F^{\mu \lambda }\textswab{J}_{ext,\lambda }\;.
\label{T_munucons}
\end{eqnarray}
Here, 
\begin{equation}
\textswab{J}_{f}^{\mu } =\textswab{n}_{f}u^{\mu }+\textswab{V}_{f}^{\mu }  \label{J_f_mu}
\end{equation}
is the electric-charge four-current of the fluid, where $\textswab{n}_{f}=u_{\nu }\textswab{J}_{f}^{\nu }$ 
is the electric-charge density and $\textswab{V}_{f}^{\mu }=\Delta _{\nu }^{\mu }\textswab{J}_{f}^{\nu }$ 
is the electric-charge diffusion current. The electric-charge four-current is related to the particle four-current
$N_f^\mu$ by $\textswab{J}_{f}^{\mu } \equiv \textswab{q} N_f^\mu$. Similarly,
$\textswab{n}_{f} = \textswab{q} n_f$, where $n_f$ is the particle density in the fluid, and
$ \textswab{V}_{f}^{\mu }=\textswab{q} V_f^\mu$, where $V_f^\mu$ is the particle diffusion current.
For the sake of generality, we have also added a source term from an external charge current
$\textswab{J}^{\mu}_{ext}$ in the energy-momentum equation (\ref{T_munucons}).

The total energy-momentum tensor of the system is given by
\begin{equation}
T^{\mu \nu }\equiv T_{em}^{\mu \nu }+T_{f}^{\mu \nu }\;.  \label{T_munu_full}
\end{equation}
It consists of an electromagnetic contribution which,
for non-polarizable, non-magnetizable fluids, reads \cite{Cercignani_book,Israel:1978up,Barrow:2006ch} 
\begin{equation}
T_{em}^{\mu \nu }\equiv -F^{\mu \lambda }F_{\left. {}\right. \lambda }^{\nu}
+\frac{1}{4}g^{\mu \nu }F^{\alpha \beta }F_{\alpha \beta }\;.
\label{T_munu_em}
\end{equation}
Here,
\begin{equation}
F^{\mu \nu }\equiv E^{\mu }u^{\nu }-E^{\nu }u^{\mu }+\epsilon ^{\mu \nu
\alpha \beta }u_{\alpha }B_{\beta }\;,  \label{F_munu}
\end{equation}
is the Faraday tensor, which we have decomposed in terms of the fluid four-velocity $u^\mu$, 
as well as the electric and magnetic field four-vectors 
$E^{\mu }\equiv F^{\mu \nu }u_{\nu }$ and $B^{\mu }\equiv 
\frac{1}{2}\epsilon ^{\mu \nu \alpha \beta }F_{\alpha \beta }u_{\nu }$, respectively,
with $\epsilon ^{\mu \nu \alpha \beta }$ being the Levi-Civita tensor.

The second part of the energy-momentum tensor (\ref{T_munu_full}) is the contribution from the fluid.
For a non-polarizable, non-magnetizable fluid it reads
\begin{equation}
T_{f}^{\mu \nu }\equiv \varepsilon u^{\mu }u^{\nu }-P\Delta ^{\mu \nu}
+2W^{\left( \mu \right. }u^{\left. \nu \right) }+\pi ^{\mu \nu }\;,
\label{T_munu_f}
\end{equation}
where we defined the energy density $\varepsilon \equiv T_{f}^{\mu \nu}u_{\mu }u_{\nu }$, the
isotropic pressure $P\equiv -\frac{1}{3}T_{f}^{\mu \nu }\Delta _{\mu \nu }$,
the energy-momentum diffusion current $W^{\mu }\equiv \Delta _{\alpha }^{\mu}T_{f}^{\alpha \beta }u_{\beta }$, 
and the shear-stress tensor $\pi ^{\mu\nu }\equiv \Delta _{\alpha \beta }^{\mu \nu }T_{f}^{\alpha \beta}$.

Maxwell's equations read \cite{degroot} 
\begin{equation}
\partial _{\mu }F^{\mu \nu }=\textswab{J}^{\nu }\;,\; \; 
\epsilon^{\mu \nu \alpha \beta} \partial _{\mu }F_{\alpha\beta }=0\;,  
\label{Maxwell_inhom_hom}
\end{equation}
where $\textswab{J}^{\mu } \equiv \textswab{J}_{f}^{\mu } + \textswab{J}_{ext}^{\mu } $ 
is the total electric charge four-current. These equations imply that
\begin{equation}
\partial _{\nu}T_{em}^{\mu \nu }=-F^{\mu \lambda }\textswab{J}_{\lambda }\;.
\end{equation}
From this and Eq.\ (\ref{T_munucons}) follows that the
energy-momentum tensor of the fluid satisfies \cite{Eckart:1940te}
\begin{equation}
\partial _{\nu }T_{f}^{\mu \nu }=F^{\mu \lambda }\textswab{J}_{f,\lambda }\;.
\label{d_mu_T_munu_f}
\end{equation}

\section{Equations of motion for the irreducible moments}
\label{sec:diss_eom_moments}

The relativistic Boltzmann equation coupled to an electromagnetic field 
\cite{deGroot,Cercignani_book}, the so-called Boltzmann-Vlasov equation
reads, 
\begin{equation}
k^{\mu }\partial _{\mu }f_{\mathbf{k}}
+\textswab{q}F^{\mu \nu }k_{\nu }\frac{\partial }{\partial k^{\mu }}f_{\mathbf{k}}=C\left[ f\right] \;,
\label{BTE_Fmunu}
\end{equation}
where $f_{\mathbf{k}}$ is the single-particle distribution function, 
$C[f]$ is the usual collision term in the Boltzmann equation, see e.g.\ Eq.\ (54) of Ref.\ \cite{Denicol:2018rbw}.

A state of local thermal equilibrium is specified by a single-particle
distribution function of the form \cite{Juttner} 
\begin{equation}
f_{0\mathbf{k}}=\left[ \exp \left( \beta_{0}E_{\mathbf{k}}-\alpha _{0}\right) +a\right] ^{-1}\;,  \label{f_0k}
\end{equation}
with $\alpha _{0}=\mu\beta _{0}$, where $\mu$ is the (in general space-time dependent) chemical potential
and $\beta _{0}=1/T$ the (space-time dependent) inverse temperature, while $a=\pm 1$ for fermions/bosons 
and $a\rightarrow 0$ for Boltzmann particles.

Unless $\alpha_0$, $\beta_0$, and $u^\mu$ are constants (i.e., independent of space-time coordinates, such that
equilibrium is global instead of local),
the distribution function $f_{0\mathbf{k}}$ is not a solution of the Boltzmann equation (\ref{BTE_Fmunu}).
However, it is a convenient starting point to derive the equations of motion for dissipative fluid dynamics using
the method of moments \cite{Denicol:2012cn,Denicol:2012es}. To this end, one decomposes
\begin{equation}
f_{\mathbf{k}}=f_{0\mathbf{k}}+\delta f_{\mathbf{k}}\;,
\label{kinetic:f=f0+df}
\end{equation}
where $\delta f_{\mathbf{k}}$ is the deviation of the solution $f_{\mathbf{k}}$ of the
Boltzmann equation from the local-equilibrium distribution function $f_{0\mathbf{k}}$.
In the following, we will use the notation
\begin{equation}
\left\langle \cdots \right\rangle \equiv \int dK \cdots
f_{\mathbf{k}}\;, \:\;\;\left\langle \cdots \right\rangle _0 \equiv \int dK \cdots
f_{0\mathbf{k}}\;, \:\;\; \left\langle \cdots \right\rangle _{\delta }\equiv \int dK\cdots
\delta f_{\mathbf{k}}\;,
\end{equation}
where $dK\equiv g\,d^{3}\mathbf{k}/[(2\pi )^{3}k^{0}]$ is the Lorentz-invariant measure in momentum space and 
$g$ is the degeneracy factor of the state with momentum $\mathbf{k}$.
From Eq.\ (\ref{kinetic:f=f0+df}) follows immediately that $\left\langle \cdots \right\rangle =
\left\langle \cdots \right\rangle_0 + \left\langle \cdots \right\rangle_\delta$.

The particle four-current and the energy-momentum tensor of the fluid are given as the following moments of $f_{\mathbf{k}}$,
\begin{equation}
N_f^\mu \equiv \left\langle k^\mu \right\rangle\;, \;\:\;
T_f^{\mu \nu} \equiv \left\langle k^\mu k^\nu\right\rangle\;,
\end{equation}
and, consequently, we identify the fluid-dynamical variables introduced in the previous section as, $n_f = \left\langle E_{\mathbf{k}} \right\rangle,\, V_f^\mu = \left\langle k^{\langle \mu \rangle} 
\right\rangle,\, \varepsilon =  \left\langle E_{\mathbf{k}}^2 \right\rangle,\,
P = -\frac{1}{3} \left\langle \Delta^{\mu \nu} k_\mu k_\nu \right\rangle,\, W^\mu =  \left\langle 
E_{\mathbf{k}} k^{\langle \mu \rangle} \right\rangle,\, \pi^{\mu \nu} =  \left\langle k^{\langle \mu} k^{\nu \rangle}
\right\rangle$. For reasons of symmetry, 
$\left\langle E_{\mathbf{k}}^{r} k^{\left\langle \mu _{1}\right. }\cdots 
k^{\left. \mu _{n}\right\rangle}\right\rangle_0$ $\equiv 0$ for $n \geq 1$, thus
$V_f^\mu = \left\langle k^{\langle \mu \rangle} \right\rangle_\delta$,
$W^\mu =  \left\langle E_{\mathbf{k}} k^{\langle \mu \rangle} \right\rangle_\delta$,
$ \pi^{\mu \nu} =  \left\langle k^{\langle \mu} k^{\nu \rangle}\right\rangle_\delta$.

Now, following Refs.\ \cite{Denicol:2012cn,Denicol:2012es} we define the
symmetric and traceless irreducible moments of $\delta f_{\mathbf{k}}$, 
\begin{equation}
\rho _{r}^{\mu _{1}\cdots \mu _{n}}\equiv \left\langle E_{\mathbf{k}}^{r}k^{\left\langle \mu _{1}\right. }\cdots 
k^{\left. \mu _{n}\right\rangle}\right\rangle _{\delta }\;.  \label{rho_r_general}
\end{equation}
Note that the tensors $k^{\left\langle \mu _{1}\right. }\cdots k^{\left. \mu _{\ell }\right\rangle}$ 
are irreducible with respect to Lorentz transformations that leave the fluid 4-velocity invariant and
form a complete and orthogonal set \cite{deGroot}. In terms of the irreducible moments (\ref{rho_r_general}) 
the corrections to the equilibrium
values of particle density, $n_{f0}$, energy density, $\varepsilon_0$, and isotropic pressure, $P_0$, are
$\delta n_{f}\equiv n_f - n_{f0} = \rho _{1}$, $\delta \varepsilon \equiv 
\varepsilon- \varepsilon_0 = \rho_{2}$, and 
$\Pi \equiv P - P_0 = (\rho _{2}- m_{0}^{2} \rho _{0})/3$. The particle
and energy-momentum diffusion currents orthogonal to the fluid velocity are 
$V_{f}^{\mu }= \rho _{0}^{\mu }$ and $W^{\mu } = \rho _{1}^{\mu }$,
while the shear-stress tensor is $\pi ^{\mu \nu } = \rho _{0}^{\mu \nu }$. 

So far, the local equilibrium state introduced in Eq.\ (\ref{f_0k}) has not been defined: the equilibrium variables 
$\alpha_{0}$, $\beta_{0}$, and $u^\mu$ must be properly specified in the context of the Boltzmann equation. 
The first step is to define temperature and chemical potential by introducing matching conditions, 
$n_f=n_{f0}(\alpha_{0},\beta_{0})$ and $\varepsilon =\varepsilon _{0}(\alpha_{0},\beta_{0})$. 
These conditions define $\alpha_{0}$ and $\beta_{0}$ such 
that the particle density and energy density of the system are identical to those of a local equilibrium state characterized 
by $f_{0\mathbf{k}}$. This implies $\rho_{1}=\rho _{2}=0$. For the sake of completeness, we shall continue with the 
derivation of the equations of motion for the irreducible moments without specifying the fluid four-velocity. 
In this way, the equations of motion derived in this paper can be made compatible with any definition of $u^\mu$. 

Equations (\ref{J_f_mucons}) and (\ref{d_mu_T_munu_f}) lead to equations of motion 
for $\alpha _{0}$, $\beta _{0}$, and $u^{\mu }$: 
\begin{eqnarray}
\dot{\alpha}_{0} &=&\frac{1}{D_{20}}\left[ -J_{30}\left( n_{f0}\theta
+\partial _{\mu }V_{f}^{\mu }\right) +J_{20}\left( \varepsilon
_{0}+P_{0}+\Pi \right) \theta +J_{20}\left( \partial _{\mu }W^{\mu }-W^{\mu }
\dot{u}_{\mu }-\pi ^{\mu \nu }\sigma _{\mu \nu }\right) +J_{20}\textswab{q}
E^{\mu }V_{f,\mu }\right] \;,  \label{D_alpha} \\
\dot{\beta}_{0} &=&\frac{1}{D_{20}}\left[ -J_{20}\left( n_{f0}\theta
+\partial _{\mu }V_{f}^{\mu }\right) +J_{10}\left( \varepsilon
_{0}+P_{0}+\Pi \right) \theta +J_{10}\left( \partial _{\mu }W^{\mu }-W^{\mu }
\dot{u}_{\mu }-\pi ^{\mu \nu }\sigma _{\mu \nu }\right) +J_{10}\textswab{q}
E^{\mu }V_{f,\mu }\right] \;,  \label{D_beta}
\end{eqnarray}
and 
\begin{align}
\dot{u}^{\mu }& =\frac{1}{\varepsilon _{0}+P_{0}}\left[ \frac{n_{f0}}{\beta_{0}}\left( \nabla ^{\mu }\alpha _{0}
-h_{0}\nabla ^{\mu }\beta _{0}\right)
-\Pi \dot{u}^{\mu }+\nabla ^{\mu }\Pi -\frac{4}{3}W^{\mu }\theta -W_{\nu}\left( \sigma ^{\mu \nu }
-\omega ^{\mu \nu }\right) -\dot{W}^{\mu }-\Delta_{\nu }^{\mu }\partial _{\kappa }\pi ^{\kappa \nu }\right]  \notag \\
& +\frac{1}{\varepsilon _{0}+P_{0}}\left[ \textswab{q}n_{f0}E^{\mu }-
\textswab{q}B\,b^{\mu \nu }V_{f,\nu }\right] \;,  \label{D_u_mu}
\end{align}
where $h_{0}\equiv \left( \varepsilon _{0}+P_{0}\right) /n_{f0}$ is the
enthalpy per particle in equilibrium and the thermodynamic integrals $J_{nq}$ and $D_{nq}$ 
are defined in App.\ \ref{kinetic_stuff}. Note that these
equations extend Eqs.\ (70) -- (72) of Ref.\ \cite{Denicol:2018rbw} by 
terms proportional to the electric field $E^{\mu }$\footnote{Note that terms proportional to $W^{\mu }$ also did 
not appear in Ref.\ \cite{Denicol:2018rbw}, since the equations derived in that reference employed the Landau 
frame~\cite{Landau_book}, where
$u^\mu$ is defined as an eigenvector of the energy-momentum tensor, $u_\mu T^{\mu \nu} = \varepsilon u^\nu$, 
leading to $W^\mu = 0$.}.

For a given fluid four-velocity $u^\mu$,
the equations of motion (\ref{J_f_mucons}) and (\ref{d_mu_T_munu_f}) only specify five
of the 14 independent variables $\alpha_0, \beta_0, \Pi, V_f^\mu, W^\mu$, and $\pi^{\mu \nu}$.
In order to close the system of equations of motion, one needs to specify additional
equations of motion that can be provided by a suitable truncation
of the infinite set of equations of motion for the irreducible moments $\rho_r^{\mu_1 \cdots \mu_\ell}$.
The latter equations are obtained by calculating the comoving derivative 
$\dot{\rho}_{r}^{\left\langle \mu _{1}\cdots \mu _{\ell }\right\rangle }
\equiv \Delta_{\nu _{1}\cdots \nu _{\ell }}^{\mu _{1}\cdots \mu _{\ell }}
u^{\alpha}\partial _{\alpha }\rho _{r}^{\nu _{1}\cdots \nu _{\ell }} $, using the
Boltzmann equation (\ref{BTE_Fmunu}), for details see Refs.\ \cite{Denicol:2012cn,Denicol:2012es,Denicol:2018rbw}.
For the irreducible moments of tensor-rank zero one obtains
\begin{align}
\dot{\rho}_{r}-C_{r-1}& =\alpha _{r}^{\left( 0\right) }\theta 
+\frac{G_{3r}}{D_{20}}\partial _{\mu }V_{f}^{\mu }-\frac{G_{2r}}{D_{20}}\partial _{\mu}W^{\mu }
+\frac{\theta }{3}\left[ m_{0}^{2}(r-1)\rho _{r-2}-(r+2)\rho _{r}-3 \frac{G_{2r}}{D_{20}}\Pi \right]  \notag \\
& +\left( r\rho _{r-1}^{\mu }+\frac{G_{2r}}{D_{20}}W^{\mu }\right) \dot{u}_{\mu }
-\nabla _{\mu }\rho _{r-1}^{\mu }+\left[ (r-1)\rho _{r-2}^{\mu \nu }+
\frac{G_{2r}}{D_{20}}\pi ^{\mu \nu }\right] \sigma _{\mu \nu }  \notag \\
& -\frac{G_{2r}}{D_{20}}\textswab{q}E_{\nu }V_{f}^{\nu }-\left( r-1\right) 
\textswab{q}E_{\nu }\rho _{r-2}^{\nu }\;,  \label{D_rho}
\end{align}
This equation is very similar to Eq.\ (75) of Ref.\ \cite{Denicol:2018rbw} 
except for the terms proportional to $W^{\mu }$ and the last two terms which constitute the contributions 
from the electric field. 

The equation of motion for irreducible moments of tensor-rank one reads
\begin{align}
\dot{\rho}_{r}^{\left\langle \mu \right\rangle }-C_{r-1}^{\left\langle \mu
\right\rangle }& =\alpha _{r}^{\left( 1\right) }\nabla ^{\mu }\alpha_{0}-\alpha _{r}^{h}\dot{W}^{\mu }
+r\rho _{r-1}^{\mu \nu }\dot{u}_{\nu }-
\frac{1}{3}\nabla ^{\mu }\left( m_{0}^{2}\rho _{r-1}-\rho _{r+1}\right)
-\Delta _{\alpha }^{\mu }\left( \nabla _{\nu }\rho _{r-1}^{\alpha \nu}
+\alpha _{r}^{h}\partial _{\kappa }\pi ^{\kappa \alpha }\right)  \notag \\
& +\frac{1}{3}\left[ m_{0}^{2}\left( r-1\right) \rho _{r-2}^{\mu }-\left(r+3\right) \rho _{r}^{\mu }
-4\alpha _{r}^{h}W^{\mu }\right] \theta +\left(r-1\right) \rho _{r-2}^{\mu \nu \lambda }\sigma _{\nu \lambda }  \notag \\
& +\frac{1}{5}\sigma ^{\mu \nu }\left[ 2 m_{0}^{2}\left(r-1\right) \rho_{r-2,\nu }
-\left( 2r+3\right) \rho _{r,\nu }-5\alpha _{r}^{h}W_{\nu }\right]
+\left( \rho _{r,\nu }+\alpha _{r}^{h}W_{\nu }\right) \omega ^{\mu \nu } 
\notag \\
& +\frac{1}{3}\left[ m_{0}^{2}r\rho _{r-1}-\left( r+3\right) \rho_{r+1}
-3\alpha _{r}^{h}\Pi \right] \dot{u}^{\mu }+\alpha _{r}^{h}\nabla^{\mu }\Pi 
-\alpha _{r}^{h}\,\textswab{q}Bb^{\mu \nu }V_{f,\nu }-\textswab{q}
Bb^{\mu \nu }\rho _{r-1,\nu }  \notag \\
& +\left( \alpha _{r}^{h}n_{f0}+\beta _{0}J_{r+1,1}\right) \textswab{q}
E^{\mu }+\frac{1}{3}\left[ \left( r+2\right) \rho _{r}-m_{0}^{2}\left(r-1\right) \rho _{r-2}\right] 
\textswab{q}E^{\mu }-\left( r-1\right) \rho_{r-2}^{\mu \nu }\textswab{q}E_{\nu }\;.  \label{D_rho_mu}
\end{align}
Here we introduced a new
dimensionless antisymmetric tensor $b^{\mu \nu }\equiv -\epsilon ^{\mu \nu \alpha \beta }u_{\alpha }b_{\beta }$, 
where the unit four-vector in the direction of the magnetic field
and orthogonal to $u^{\mu }$ is $b^{\mu }\equiv \frac{B^{\mu }}{B}$, with 
$b^{\mu }b_{\mu }=-1$ and $B \equiv \sqrt{-B^\mu B_\mu}$. 
Equation (\ref{D_rho_mu}) differs from Eq.\ (76) of Ref.\ \cite{Denicol:2018rbw} by
the last three terms taking into account the electric field, as well as by
the additional terms proportional to $W^{\mu }$. 

Finally, the equation of motion for the irreducible moments of tensor-rank two is 
\begin{align}
\dot{\rho}_{r}^{\left\langle \mu \nu \right\rangle }-C_{r-1}^{\left\langle
\mu \nu \right\rangle }& =2\alpha _{r}^{\left( 2\right) }\sigma ^{\mu \nu }+
\frac{2}{15}\left[ m_{0}^{4}\left( r-1\right) \rho _{r-2}-m_{0}^{2}\left(2r+3\right) \rho _{r}
+\left( r+4\right) \rho _{r+2}\right] \sigma ^{\mu \nu}+\frac{2}{5}\dot{u}^{\left\langle \mu \right. }
\left[ m_{0}^{2}r\rho_{r-1}^{\left. \nu \right\rangle }-\left( r+5\right) \rho _{r+1}^{\left. \nu
\right\rangle }\right]  \notag \\
& -\frac{2}{5}\left[ \nabla ^{\left\langle \mu \right. }\left( m_{0}^{2}\rho_{r-1}^{\left. \nu \right\rangle }
-\rho _{r+1}^{\left. \nu \right\rangle}\right) \right] +r\rho _{r-1}^{\mu \nu \gamma }\dot{u}_{\gamma }
-\Delta_{\alpha \beta }^{\mu \nu }\nabla _{\lambda }\rho _{r-1}^{\alpha \beta\lambda }
+\left( r-1\right) \rho _{r-2}^{\mu \nu \lambda \kappa }\sigma_{\lambda \kappa }
+2\rho _{r}^{\lambda \left\langle \mu \right. }\omega_{\left. {}\right. \lambda }^{\left. \nu \right\rangle }  \notag \\
& +\frac{1}{3}\left[ m_{0}^{2}\left( r-1\right) \rho _{r-2}^{\mu \nu}-\left( r+4\right) \rho _{r}^{\mu \nu }\right] \theta 
+\frac{2}{7}\left[2m_{0}^{2}\left( r-1\right) \rho _{r-2}^{\kappa \left\langle \mu \right.}
-\left( 2r+5\right) \rho _{r}^{\kappa \left\langle \mu \right. }\right]
\sigma _{\kappa }^{\left. \nu \right\rangle }-2\,\textswab{q}Bb^{\alpha\beta }\Delta _{\alpha \kappa }^{\mu \nu }
g_{\lambda \beta }\rho_{r-1}^{\kappa \lambda }  \notag \\
& +2\textswab{q}E^{\left\langle \mu \right. }\rho _{r}^{\left. \nu\right\rangle }
-\left( r-1\right) \Delta _{\alpha \beta }^{\mu \nu }\left[ \textswab{q}E_{\lambda }\rho _{r-2}^{\alpha \beta \lambda }
+\frac{2}{5} \textswab{q}E^{\left( \alpha \right. }\left( m_{0}^{2}\rho _{r-2}^{\left.\beta \right) }
-\rho _{r}^{\left. \beta \right) }\right) \right] \;.
\label{D_rho_munu}
\end{align}
This equation differs from Eq.\ (77) of Ref.\ \cite{Denicol:2018rbw} by the last two terms, which constitute
the contributions from a non-vanishing electric field.

In Eqs.\ (\ref{D_rho}), (\ref{D_rho_mu}), and (\ref{D_rho_munu}), $\alpha _{r}^{h}$, $\alpha _{r}^{\left( \ell \right) }$,
and $G_{ij}$ are thermodynamic coefficients, which are explicitly given in App.\ \ref{kinetic_stuff}, while the
linearized collision integral is defined as 
\begin{eqnarray}
\mathcal{C}_{r-1}^{\left\langle \mu _{1}\cdots \mu _{\ell }\right\rangle }
&\equiv &\Delta _{\nu _{1}\cdots \nu _{\ell }}^{\mu _{1}\cdots \mu _{\ell}}
\int dK\,E_{\mathbf{k}}^{r-1}\,k^{\nu _{1}}\cdots k^{\nu _{\ell }}C\left[ f\right]  
=-\sum_{n=0}^{N_{\ell }}\mathcal{A}_{rn}^{\left( \ell \right) }\rho_{n}^{\mu _{1}\cdots \mu _{\ell }}\;,  \label{Lin_collint}
\end{eqnarray}
where the coefficient $\mathcal{A}_{rn}^{\left( \ell \right) }\sim \lambda _{\mathrm{mfp}}$ contains time scales 
proportional to the mean free path of the particles. Note that the last equality of the above equation is obtained using 
the moment expansion of the single-particle distribution function first introduced in Ref.\ \cite{Denicol:2012cn}, 
which, for the sake of completeness, is listed in App.\ \ref{kinetic_stuff}.

\section{Equations of motion in the 14-moment approximation}

\label{sec:diss_eom_2ndorder}

In order to obtain a closed system of fluid-dynamical equations of motion, we now 
truncate the infinite set (\ref{D_rho}) -- (\ref{D_rho_munu}) of equations of motion 
for the irreducible moments. The simplest and most widely used truncation is the so-called 14-moment 
approximation \cite{Israel:1979wp}.
First, all  irreducible tensor moments $\rho_{r}^{\mu _{1}\cdots \mu _{\ell }}$ for $\ell >2$ are explicitly set to
zero in Eqs.\ (\ref{D_rho_mu}) -- (\ref{D_rho_munu}). Second,
the remaining scalar $\rho _{r}$, vector $\rho _{r}^{\mu }$, and
rank-2 tensor moments $\rho _{r}^{\mu \nu }$ are expressed as linear combinations
of the lowest-order moments $\rho _{0} \equiv -3\Pi /m_{0}^{2}$, $\rho _{0}^{\mu }\equiv V_f^{\mu }$, 
$\rho_{1}^{\mu }\equiv W^{\mu }$, and $\rho _{0}^{\mu \nu }\equiv \pi ^{\mu \nu }$, i.e., in terms of
quantities appearing in $\textswab{J}_f^\mu$ and $T_f^{\mu \nu}$, cf.\ Eqs.\ (\ref{J_f_mu}), (\ref{T_munu_f}).
The relations affecting this truncation are Eqs.\ (\ref{OMG_rho}) -- (\ref{OMG_rho_mu_nu}).

Equation (\ref{D_rho}) then leads to an equation of motion for the bulk viscous pressure 
\begin{align}
\tau _{\Pi }\dot{\Pi}+\Pi & =-\zeta \theta -\delta _{\Pi \Pi }\,\Pi \theta
+\lambda _{\Pi \pi }\,\pi ^{\mu \nu }\sigma _{\mu \nu }  
-\ell _{\Pi V}\,\nabla _{\mu }V_{f}^{\mu }-\tau _{\Pi V}\,V_{f}^{\mu }\dot{u}_{\mu }
-\lambda _{\Pi V}\,V_{f}^{\mu }\nabla _{\mu }\alpha _{0}  \notag \\
& -\ell _{\Pi W}\,\nabla _{\mu }W^{\mu }-\tau _{\Pi W}\,W^{\mu }\dot{u}_{\mu}
-\lambda _{\Pi W}\,W^{\mu }\nabla _{\mu }\alpha _{0}  
 -\delta _{\Pi VE}\textswab{q}V_{f}^{\nu }E_{\nu }-\delta _{\Pi WE}
\textswab{q}W^{\nu }E_{\nu }\;.  \label{relax_bulk}
\end{align}
Similarly, from Eq.\ (\ref{D_rho_mu}) we obtain an equation for the particle- and energy-diffusion currents,
\begin{align}
 \tau _{V}\dot{V}_{f}^{\left\langle \mu \right\rangle }-\tau _{V}h_{0}^{-1}
\dot{W}^{\left\langle \mu \right\rangle} & +V_{f}^{\mu }-h_{0}^{-1}W^{\mu }  
 =\kappa \nabla ^{\mu }\alpha _{0}-\tau _{V}V_{f,\nu }\omega ^{\nu \mu}-\delta _{VV}\,V_{f}^{\mu }\theta
  -\lambda _{VV}\,V_{f,\nu }\sigma ^{\mu\nu }  \notag \\
& +\tau _{V}h_{0}^{-1}W_{\nu }\omega ^{\nu \mu }-\delta _{WW}\,W^{\mu}\theta 
-\lambda _{WW}\,W_{\nu }\sigma ^{\mu \nu }   -\ell _{V\Pi }\nabla ^{\mu }\Pi 
+\ell _{V\pi }\Delta ^{\mu \nu }\nabla_{\lambda }\pi _{\nu }^{\lambda }+\tau _{V\Pi }\,\Pi \dot{u}^{\mu }  \notag
\\
& -\tau _{V\pi }\,\pi ^{\mu \nu }\dot{u}_{\nu }+\lambda _{V\Pi }\,\Pi \nabla^{\mu }\alpha _{0}
-\lambda _{V\pi }\,\pi ^{\mu \nu }\nabla _{\nu }\alpha _{0}
 -\delta _{VB}\,\textswab{q}Bb^{\mu \nu }V_{f,\nu }-\delta _{WB}\,
\textswab{q}Bb^{\mu \nu }W_{\nu }  \notag \\
& +\delta _{VE}\textswab{q}E^{\mu }+\delta _{V\Pi E}\textswab{q}\Pi E^{\mu}
+\delta _{V\pi E}\textswab{q}\pi ^{\mu \nu }E_{\nu }\;,  \label{relax_heat}
\end{align}
The equation of motion for the shear-stress tensor follows from Eq.\ (\ref{D_rho_munu}), 
\begin{align}
\tau _{\pi }\dot{\pi}^{\left\langle \mu \nu \right\rangle }+\pi ^{\mu \nu }&
=2\eta \sigma ^{\mu \nu }+2\tau _{\pi }\pi _{\lambda }^{\left\langle \mu\right. }\omega ^{\left. \nu \right\rangle \lambda }
-\delta _{\pi \pi }\,\pi^{\mu \nu }\theta -\tau _{\pi \pi }\,\pi ^{\lambda \left\langle \mu \right.}
\sigma _{\lambda }^{\left. \nu \right\rangle }+\lambda _{\pi \Pi }\,\Pi \sigma ^{\mu \nu }  \notag \\
& -\tau _{\pi V}\,V_{f}^{\left\langle \mu \right. }\dot{u}^{\left. \nu \right\rangle }
+\ell _{\pi V}\nabla ^{\left\langle \mu \right.}V_{f}^{\left. \nu \right\rangle }
+\lambda _{\pi V}\,V_{f}^{\left\langle \mu \right. }\nabla ^{\left. \nu \right\rangle }\alpha _{0}  
-\tau _{\pi W}\,W^{\left\langle \mu \right. }\dot{u}^{\left. \nu \right\rangle }
+\ell _{\pi W}\nabla ^{\left\langle \mu \right. }W^{\left.\nu \right\rangle }
+\lambda _{\pi W}\,W^{\left\langle \mu \right. }\nabla^{\left. \nu \right\rangle }\alpha _{0}  \notag \\
& -\delta _{\pi B}\,\textswab{q}Bb^{\alpha \beta }\Delta _{\alpha \kappa}^{\mu \nu }
g_{\lambda \beta }\pi ^{\kappa \lambda }   
+\delta _{\pi VE}\textswab{q}E^{\left\langle \mu \right. }V_{f}^{\left.\nu \right\rangle }
+\delta _{\pi WE}\textswab{q}E^{\left\langle \mu \right.}W^{\left. \nu \right\rangle }\; .  \label{relax_shear}
\end{align}
The coefficients appearing in these equations are listed in App.\ \ref{AppB}.

Note that Eq.\ (\ref{relax_heat}) represents the relaxation equation for the
heat flow defined by
\begin{equation}
q^{\mu } \equiv W^{\mu }-h_{0}V_{f}^{\mu }\;.
\end{equation}
In case we choose the local rest frame following Landau's picture 
(which imposes $W^{\mu }\equiv 0$), the heat flow is simply given in terms of the particle
diffusion alone, $q^{\mu } = -h_{0}V_{f}^{\mu }$. On the other hand,
choosing the rest frame according to Eckart's picture (which requires $V_{f}^{\mu }\equiv 0$), 
leads to a heat flow that is solely given by the flow of
energy and momentum, $q^{\mu } = W^{\mu }$. Since the relaxation equations 
(\ref{relax_bulk}) -- (\ref{relax_shear}) contain both diffusive quantities, the equations of motion derived in this paper
are consistent with either choice of local rest frame.

The coefficients proportional to the electric field in the 
equation for the bulk viscous pressure are
\begin{equation}
\delta _{\Pi VE} =\frac{m_{0}^{2}}{3\mathcal{A}_{00}^{\left(0\right) }}\left( \mathcal{F}_{20}^{(1)} 
-\frac{G_{20}}{D_{20}} 
- \frac{\beta_0}{h_0} \frac{\partial \mathcal{F}_{10}^{(1)}}{\partial \beta_0}\right)\;, \;\;
\delta _{\Pi WE} =\frac{m_{0}^{2}}{3\mathcal{A}_{00}^{\left(0\right) }}
\left( \mathcal{F}_{21}^{(1)} - \frac{\beta_0}{h_0} \frac{\partial \mathcal{F}_{11}^{(1)}}{\partial \beta_0} \right)\; .
\end{equation}
The coefficients proportional to the electric field in the equation for the particle-diffusion current are
\begin{eqnarray}
\delta _{VE} &=&\frac{1}{\mathcal{A}_{00}^{\left( 1\right) }}\left(-n_{f0} h^{-1}_{0}+\beta _{0}J_{11}\right) \; , \\ 
\delta _{V\Pi E} &=&-\frac{1}{m_{0}^{2}\mathcal{A}_{00}^{\left(1\right) }}
\left( 2 + m_{0}^{2}\mathcal{F}_{20}^{(1)} 
- m_0^2\frac{\beta_0}{h_0} \frac{\partial \mathcal{F}_{10}^{(0)}}{\partial \beta_0}  \right) \; , \;\;
\delta _{V\pi E} = \frac{1}{\mathcal{A}_{00}^{\left(1\right) }} 
\left(\mathcal{F}_{20}^{(2)}- \frac{\beta_0}{h_0} \frac{\partial \mathcal{F}_{10}^{(2)}}{\partial \beta_0} \right)\;,
\end{eqnarray}
and the coefficient coupling $W^\mu$ to the magnetic field is
\begin{equation}
\delta _{WB}=\frac{\mathcal{F}_{11}^{(1)}}{\mathcal{A}_{00}^{\left( 1\right) }}\;.
\end{equation}
Finally, the coefficients proportional to the electric field in the equation for the shear-stress tensor are
\begin{equation}
\delta _{\pi VE} =\frac{2}{5 \mathcal{A}_{00}^{\left( 2\right) }}
\left( 4+ m_{0}^{2} \mathcal{F}_{20}^{(1)} 
-m_{0}^{2} \frac{\beta_0}{h_0} \frac{\partial \mathcal{F}_{10}^{(1)}}{\partial \beta_0} \right) \;, \;\;\;
\delta _{\pi WE} =\frac{2 m_0^2}{5 \mathcal{A}_{00}^{\left( 2\right) }}
\left( \mathcal{F}_{21}^{(1)} - \frac{\beta_0}{h_0} 
\frac{\partial \mathcal{F}_{11}^{(1)}}{\partial \beta_0} \right)\; .
\end{equation}
The thermodynamic integral $ \mathcal{F}_{rn}^{(\ell)}$ is defined in Eq.\ (\ref{F_rn}). 

In the limit of a massless Boltzmann gas with constant cross section $\sigma$,
$J_{nq}\equiv I_{nq}=\frac{\left( n+1\right) !}{2\left( 2q+1\right) !!}\beta_{0}^{2-n}P_{0}$, and hence 
$\mathcal{A}_{00}^{(1)}=4/(9\lambda _{\mathrm{mfp}})$, 
$\mathcal{A}_{00}^{(2)}=3/(5\lambda _{\mathrm{mfp}})$, where $\lambda _{\mathrm{mfp}}=1/(n_{0}\sigma )$ 
is the mean free path of the
particles. In the massless limit, $m_0=0$, the coefficients $\delta _{\Pi VE}=\delta _{\Pi WE}= 0$, while
$\delta _{V\Pi E}$ formally diverges $\sim 1/m_0^2$. However, the bulk viscous pressure is $\Pi = - m_0^2\rho_0/3$, which
cancels this divergence, and the remaining term is $\sim \rho_0 E^\mu$. Now, $E^\mu$ is of order one in gradients 
[see below and Ref.\ \cite{Hernandez:2017mch}], while $\rho_0$ is actually of second order, since the coefficient
$\alpha_r^{(0)}$ in the Navier-Stokes term in Eq.\ (\ref{D_rho}) vanishes in the massless limit for all $r$. 
Thus, the respective term is of third order in gradients and, for this reason, we neglect it in the massless limit.

In Table \ref{diff_massless} we list the $m_0=0$ values of those coefficients in Eq.\ (\ref{relax_heat}), which are not
proportional to $\Pi$.
\begin{table}[h]
\begin{center}
\begin{tabular}{|c|c|c|c|c|c|c|c|c|c|c|c|c|}
\hline
$\kappa$ & $\tau_{V}[\lambda_{\mathrm{mfp}}]$ & $\delta_{VV}[\tau_{V}]$ 
& $\delta_{WW}[\tau_{V}]$ & ${\lambda}_{VV}[\tau_{V}]$ & ${\lambda}_{WW}[\tau_{V}]$ & 
${\lambda }_{V\pi }[\tau_{V}]$ & $\ell_{V\pi }[\tau_{V}]$ & $\tau_{V\pi }[\tau_{V}]$ &
$\delta_{VB }[\tau_{V}]$ & $\delta_{WB }[\tau_{V}]$ & $\delta_{VE }[\tau_{V}]$ & 
$\delta_{V\pi E }[\tau_{V}]$ \\ \hline
${3}/\left( 16{\sigma }\right) $ & $9/4$ & $1$ & $-\beta_{0}/3$ &  $3/5$ & $-\beta_{0}/4$ 
& $\beta _{0}/{20}$ & ${\beta _{0}}/{20}$ & ${\beta _{0}}/{20}$ 
& $5\beta_0/12$ & $-\beta^2_{0}/12$ & $P_0\beta^2_{0}/12$ & $0$\\ \hline
\end{tabular}%
\end{center}
\caption{{\protect\small The coefficients for the diffusion equation for a
Boltzmann gas with constant cross section in the ultrarelativistic limit, in
the 14-moment approximation, with $\tau^{(1)}_{00} = \tau_V$}.}
\label{diff_massless}
\end{table}

Similarly, in Table \ref{shear_massless} we list the $m_0=0$ values of those coefficients in Eq.\ (\ref{relax_shear}), 
which are not proportional to $\Pi$.
\begin{table}[h]
\begin{center}
\begin{tabular}{|c|c|c|c|c|c|c|c|c|c|c|c|c|}
\hline
$\eta$ & $\tau_{\pi }[\lambda_{\mathrm{mfp}}]$ & $\delta _{\pi \pi }[\tau_{\pi}] $ 
& ${\tau}_{\pi \pi }[\tau_{\pi}]$ & ${\lambda }_{\pi V}[\tau _{\pi}]$ 
& $\ell_{\pi V}[\tau _{\pi}]$ & $\tau_{\pi V}[\tau _{\pi}]$
& ${\lambda }_{\pi W}[\tau _{\pi}]$ 
& $\ell_{\pi W}[\tau _{\pi}]$ & $\tau_{\pi W}[\tau _{\pi}]$ 
& $\delta_{\pi B}[\tau _{\pi}]$ & $\delta_{\pi V E}[\tau _{\pi}]$ & $\delta_{\pi W E}[\tau _{\pi}]$ \\ \hline
${4}/({3\sigma \beta _{0}})$ & $5/3$ & $4/3$ & $10/7$  & $0$  & $0$ & $0$ 
& $0$ & $2/5$ & $2$ & $2\beta_{0}/5$ & $8/5$ & $0$ \\ 
\hline
\end{tabular}
\end{center}
\caption{{\protect\small The coefficients for the shear-stress equation for a
Boltzmann gas with constant cross section in the ultrarelativistic limit, in
the 14-moment approximation, with  $\tau^{(2)}_{00} = \tau_\pi$}.}
\label{shear_massless}
\end{table}

\section{Navier-Stokes limit, Ohmic current, and Wiedemann--Franz law}
\label{NS_theory}

In the Navier-Stokes limit, all second-order terms are discarded from
the relaxation equations (\ref{relax_bulk}) -- (\ref{relax_shear}). We employ the power-counting
advertised in Ref.\ \cite{Hernandez:2017mch}, i.e., $E^\mu$ is of order one, i.e., of the same order as gradients
of $\alpha_0$, $\beta_0$, and $u^\mu$, or of the same order as the dissipative quantities
$\Pi$, $V_f^\mu$, $W^\mu$, and $\pi^{\mu \nu}$. On the other hand, the magnetic field
is of order zero, like other thermodynamic quantities. For the
bulk viscous pressure and shear-stress tensor, this ultimately leads to $\Pi =-\zeta \theta $ and $\pi ^{\mu \nu }=2\eta
\sigma ^{\mu \nu }-\delta _{\pi B}\,\textswab{q}Bb^{\alpha \beta }\Delta_{\alpha \kappa }^{\mu \nu }
g_{\lambda \beta }\pi ^{\kappa \lambda }$, see
the discussion in Sec.\ IV.B of Ref.\ \cite{Denicol:2018rbw}, where these equations have already been analyzed.

However, for the Navier-Stokes limit of the diffusion currents, the electric field has a non-negligible impact.
For the sake of simplicity and comparison to Ref.\ \cite{Denicol:2018rbw}, we work in the Landau frame,
where $W^\mu = 0$. To first order, the particle-diffusion current becomes
\begin{equation} \label{1storderVf}
V_{f}^{\mu }=\kappa \nabla^{\mu }\alpha_{0} + \delta _{VE}\textswab{q}E^{\mu}
-\delta _{VB}\,\textswab{q}Bb^{\mu \nu }V_{f, \nu }\;.
\end{equation}
The Ohmic induction current is given by the second term of Eq.\ (\ref{1storderVf}) (after multiplying
by $\textswab{q}$), 
\begin{equation} \label{simple_induction}
\textswab{J}_{ind}^\mu \equiv \sigma_E E^\mu\;,
\end{equation}
with the electric conductivity 
\begin{equation}
\sigma_E \equiv \textswab{q}^2 \delta_{VE}\;.
\end{equation}
As originally noted by Einstein \cite{Arnold:2000dr}, the electric conductivity and the particle-diffusion coefficient
must be related by
\begin{equation} \label{Einstein}
\sigma_E = \textswab{q}^2 \beta_0 \kappa\;, 
\end{equation}
which is the kinetic-theory version of the famous Wiedemann--Franz law. For the massless Boltzmann gas,
the validity of this relation can be easily checked using the relation
$\delta_{VE} = \frac{3}{16}n_{f0}\beta_{0}\lambda_{\mathrm{mfp}}$ and
the fact that $\kappa =\frac{3}{16} n_{f0} \lambda_{\mathrm{mfp}}$ \cite{Denicol:2012cn}.
As noted in Ref.\ \cite{Kovtun:2016lfw}, this relation must also hold for a different reason: 
in a state of constant $T$ and $u^\mu$ and in the absence of dissipation,
an electric field induces a charge-density gradient such that (in our conventions for metric and 
chemical potential), 
\begin{equation} \label{Kovtun}
\nabla^\mu \alpha_0 = - \textswab{q} \beta_0 E^\mu\;.
\end{equation}
This relation can also be found from the
second-order transport equation (\ref{relax_heat}), setting all dissipative quantities to zero, which
leads to the condition
$\kappa \nabla^{\mu }\alpha_{0} = - \delta _{VE}\textswab{q}E^{\mu}$. This relation  
together with Eq.\ (\ref{Kovtun}) then confirms the Einstein relation (\ref{Einstein}).

Note that in the presence of a magnetic field Eq.\ (\ref{simple_induction}) no longer 
holds \cite{Bekenstein_78}. Using Eq.\ (\ref{D_rho_mu}) in the Navier-Stokes approximation 
we obtain
\begin{equation}
\rho _{r}^{\mu }=\kappa _{r}^{\mu \nu }\nabla _{\nu }\alpha_{0}
+\delta_{r}^{\mu \nu }\textswab{q} E_{\nu }\;,
\end{equation}
hence the conductivity tensor can be defined similarly to Eq. (\ref{Einstein})
\begin{equation}
\sigma_{E,r}^{\mu \nu} = \textswab{q}^2 \delta_{r}^{\mu \nu }\;.
\end{equation}
The rank-two tensor coefficients may be decomposed in the direction parallel and orthogonal to the magnetic field in 
terms of the projection operators, $b^{\mu }b^{\nu }$, 
$\Xi^{\mu \nu} \equiv \Delta^{\mu \nu} + b^{\mu} b^{\nu}$, and the tensor $b^{\mu \nu }$ as
\begin{eqnarray}
\kappa_{r}^{\mu \nu } &=&\kappa_{r\perp }\Xi^{\mu \nu } 
-\kappa_{r\parallel }b^{\mu }b^{\nu }-\kappa_{r\times }b^{\mu \nu }\;,
\label{kappa_NS} \\
\delta_{r}^{\mu \nu } &=&\delta_{r\perp }\Xi^{\mu \nu} 
-\delta_{r\parallel }b^{\mu }b^{\nu }-\delta_{r\times }b^{\mu \nu }\;.
\label{sigma_E_NS}
\end{eqnarray}
In order to calculate the transport coefficients $\kappa _{r}^{\mu \nu }$ or 
$\delta _{r}^{\mu \nu }$, we will follow the inversion procedure of Ref.\
\cite{Denicol:2018rbw}, hence in the 14-moment approximation ($N_{1}=1$),
setting $\nabla ^{\mu }\alpha _{0}=0$ we get
\begin{equation}
\delta_{0\parallel } 
 =  \frac{\beta_0 \alpha_r^{(1)}}{\mathcal{A}_{r0}^{\left( 1\right)}}\;,\;\;
\delta_{0\perp } =\delta_{0\parallel }\left[ 1+\left( \textswab{q}B\frac{\mathcal{F}_{1-r,0}^{\left(1\right) }
+\alpha_{r}^{h}}{\mathcal{A}_{r0}^{\left( 1\right) }}\right)^{2}\right]^{-1}\;,\;\;
\delta_{0\times }=\delta_{0\perp }\,\textswab{q}B
\frac{\mathcal{F}_{1-r,0}^{\left( 1\right) }+\alpha _{r}^{h}}
{\mathcal{A}_{r0}^{\left( 1\right) }}\;.
\end{equation}
Comparing with Eqs.\ (101) of Ref.\ \cite{Denicol:2018rbw}, 
we conclude that
\begin{equation}
\delta_{0\parallel} = \beta_0 \kappa_{0 \parallel}\;, \;\;\delta_{0\perp }=\beta_0 \kappa _{0}\;,\;\;
\delta_{0\times }= \beta_0 \kappa_{0\times } \; ,
\end{equation}
confirming that Eq.\ (\ref{Einstein}) also holds in tensorial form,
$\sigma_{E,r}^{\mu \nu} = \textswab{q}^2 \beta_0 \kappa_r^{\mu \nu}$, and irrespective
of the limit of a massless Boltzmann gas.

\section{Conclusions and outlook}

\label{conclusions}

Based on the moment expansion of the Boltzmann equation for a single-component gas of charged spin-zero 
particles coupled to an electromagnetic field, we have derived the equations of motion of resistive,
second-order dissipative magnetohydrodynamics in the 14-moment approximation. New transport 
coefficients appear due to the coupling to the electric field. We computed these coefficients in the limit
of a massless Boltzmann gas. We analyzed the Navier-Stokes limit of the dissipative quantities
and recovered Ohm's law. We found that the electrical conductivity and the particle diffusion 
satisfy the well-known Einstein relation, which constitutes a type of Wiedemann-Franz law.

In future studies, one should address the generalization to particles with non-zero spin. Then, particles
have a microscopic dipole moment which generates non-vanishing macroscopic magnetization and polarization fields
\cite{Israel:1978up,Kovtun:2016lfw}. The spin of the particles also gives rise to spin-vorticity coupling terms, which
leads to the so-called Chiral Vortical Effect \cite{Son:2009tf}.
This may necessitate an extension of the standard fluid-dynamical conservation laws 
by an equation of motion for the macroscopic spin tensor \cite{Florkowski:2017ruc,Becattini:2018duy}.

\begin{acknowledgments}

The authors acknowledge enlightening discussion with G.\ Moore.
E.M.\ acknowledges the warm hospitality of the Department of
Physics of the University of Jyv\"askyl\"a, where part of this work was done.
This work was supported by the Deutsche Forschungsgemeinschaft (DFG, German Research Foundation) 
through the Collaborative Research Center CRC-TR 211 ``Strong-interaction matter
under extreme conditions'' -- project number 315477589 - TRR 211.
G.S.D.\ thanks for Conselho Nacional de Desenvolvimento Cient\'ifico e Tecnol\'ogico (CNPq) for financial support. 
E.M.\ is supported by the Bundesministerium f\"ur Bildung und Forschung (BMBF) and by the Research 
Council of Norway, (NFR) Project No.~255253/F50.
H.N.\ is supported by the
Academy of Finland, project 297058.
D.H.R.\ is partially supported by the High-end Foreign Experts project GDW20167100136 of the State
Administration of Foreign Experts Affairs of China. 
\end{acknowledgments}

\appendix

\section{Some useful formulas}

\label{kinetic_stuff}

Following Refs.\ \cite{Denicol:2012cn,Denicol:2012es}, we recall that the
single-particle distribution function $f_{\mathbf{k}}$ can be expanded
around $f_{0\mathbf{k}}$ as,
\begin{equation}
f_{\mathbf{k}}=f_{0\mathbf{k}}+f_{0\mathbf{k}}\left( 1-af_{0\mathbf{k}}\right) 
\sum_{\ell =0}^{\infty }\sum_{n=0}^{N_{\ell }}\rho _{n}^{\mu_{1}\cdots \mu _{\ell }}
k_{\left\langle \mu _{1}\right. }\cdots k_{\left.\mu _{\ell }\right\rangle }\mathcal{H}_{\mathbf{k}n}^{(\ell )}\;.
\label{f_iso_expansion}
\end{equation}
where the coefficient $\mathcal{H}_{\mathbf{k}n}^{(\ell )}$ is a polynomial
in energy and defined as, 
\begin{equation}
\mathcal{H}_{\mathbf{k}n}^{(\ell )}=\frac{\left( -1\right)^{\ell }}{\ell ! \, J_{2\ell ,\ell }}
\sum_{i=n}^{N_{\ell }}\sum_{m=0}^{i}a_{in}^{(\ell)}a_{im}^{(\ell )}E_{\mathbf{k}}^{m}\;.  \label{H_kn}
\end{equation}
The coefficients $a_{ij}^{(\ell )}$ are calculated via the Gram-Schmidt
orthogonalization procedure and are expressed in terms of thermodynamic
integrals $J_{nq}$, for more details see for example Ref.\
\cite{Denicol:2012cn}.

Any irreducible moment of arbitrary order $r$ and tensor rank $\ell $ can
always be expressed as a linear combination of irreducible moments of all
orders $n$ and the same tensor rank, 
\begin{equation}
\rho _{r}^{\mu _{1}\cdots \mu _{\ell }}=\sum_{n=0}^{N_{\ell }}\rho _{n}^{\mu_{1}\cdots \mu _{\ell }}
\mathcal{F}_{-r,n}^{\left( \ell \right) }\;,
\label{useful}
\end{equation}
where
\begin{equation}
\mathcal{F}_{rn}^{\left( \ell \right) }=\frac{\ell !}{\left( 2\ell +1\right)!!}
\int dKE_{\mathbf{k}}^{-r}\mathcal{H}_{\mathbf{k}n}^{\left( \ell \right)}
\left( \Delta ^{\alpha \beta }k_{\alpha }k_{\beta }\right) ^{\ell }f_{0\mathbf{k}}\left( 1-af_{0\mathbf{k}}\right) \;.  \label{F_rn}
\end{equation}
In the 14-moment approximation the above expressions simplify
considerably, hence using Eq.\ (\ref{useful}) with the summation limits
$N_{0}=2,\,N_{1}=1,\,N_{2}=0$ for different tensor rank, we obtain the
following relations,
\begin{align}
\rho _{r}& \equiv \sum_{n=0,\neq 1,2}^{N_{0}}\rho _{n}\mathcal{F}_{-r,n}^{\left( 0\right) }
= -\frac{3}{m_{0}^{2}}\Pi \mathcal{F}_{-r,0}^{\left( 0\right) } 
 \equiv -\frac{3}{m_{0}^{2}}\Pi \frac{J_{r0}D_{30}+J_{r+1,0}G_{23}+J_{r+2,0}D_{20}}{
J_{20}D_{20}+J_{30}G_{12}+J_{40}D_{10}}\;,  \label{OMG_rho} \\
\rho _{r}^{\mu }& \equiv \sum_{n=0}^{N_{1}}\rho _{n}^{\mu }\mathcal{F}_{-r,n}^{\left( 1\right) }
= V_{f}^{\mu }\mathcal{F}_{-r,0}^{\left(1\right) }+W^{\mu }\mathcal{F}_{-r,1}^{\left( 1\right) } 
 \equiv V_{f}^{\mu }\frac{J_{r+2,1}J_{41}-J_{r+3,1}J_{31}}{D_{31}}+W^{\mu }
\frac{-J_{r+2,1}J_{31}+J_{r+3,1}J_{21}}{D_{31}}\;,  \label{OMG_rho_mu} \\
\rho _{r}^{\mu \nu }& \equiv \sum_{n=0}^{N_{2}}\rho _{n}^{\mu \nu }\mathcal{F}_{-r,n}^{\left( 2\right) }
= \pi ^{\mu \nu }\mathcal{F}_{-r,0}^{\left(2\right) }\equiv \pi ^{\mu \nu }\frac{J_{r+4,2}}{J_{42}}\;.
\label{OMG_rho_mu_nu}
\end{align}
Note that Eqs.\ (\ref{OMG_rho}) -- (\ref{OMG_rho_mu_nu}) are the same as Eqs.\
(115) -- (117) of Ref.\ \cite{Denicol:2018rbw}, except for Eq.\ (\ref{OMG_rho_mu}),
which now contains a term proportional to $W^{\mu }$ when compared to Eq.\
(116) of Ref.\ \cite{Denicol:2018rbw}. Furthermore, for $r,n\geq 0$, 
$\mathcal{F}_{-r,n}^{\left( \ell \right) }=\delta _{r n}$, 
however Eqs.\ (\ref{OMG_rho}) -- (\ref{OMG_rho_mu_nu}) are to
be used for irreducible moments not only with positive but also with
negative $r$ given by
\begin{equation}
\rho _{-r}^{\mu _{1}\cdots \mu _{\ell }}=\sum_{n=0}^{N_{\ell }}\rho_{n}^{\mu _{1}\cdots \mu _{\ell }}
\mathcal{F}_{rn}^{\left( \ell \right) }\;.
\label{rho_negative_r}
\end{equation}
Truncating the sum as in Eqs.\ (\ref{OMG_rho}) -- (\ref{OMG_rho_mu_nu}), 
the coefficients of Eq.\ (\ref{rho_negative_r}) can be
written similarly as in Eq.\ (67) of Ref.\ \cite{Denicol:2012cn},
\begin{align}
\rho _{-r}& =-\frac{3}{m_0^{2}}\,\gamma _{r}^{(0)}\Pi +\mathcal{O}(\mathrm{Kn})\;,\text{ \ } \\
\rho _{-r}^{\mu }& =\gamma _{r}^{V(1)}V_{f}^{\mu }+\gamma _{r}^{W(1)}W^{\mu
}+\mathcal{O}(\mathrm{Kn})\;,\text{ \ } \\
\rho _{-r}^{\mu \nu }& =\gamma _{r}^{(2)}\pi ^{\mu \nu }+\mathcal{O}(\mathrm{Kn})\;.
\end{align}
In Ref.\ \cite{Denicol:2012cn} the coefficients $\gamma_r^{(\ell)}$ were calculated 
explicitly in the Landau frame, hence $\gamma _{r}^{V(1)} \equiv \gamma _{r}^{(1)}$, while 
$\gamma _{r}^{W(1)}$ is a new coefficient in the Eckart frame. Note that, in the 14-moment approximation,
$\gamma_r^{(0)} \equiv \mathcal{F}_{r0}^{(0)}, \gamma_r^{V(1)} \equiv \mathcal{F}_{r0}^{(1)},
\gamma_r^{W(1)} \equiv \mathcal{F}_{r1}^{(1)}, \gamma_r^{(2)} \equiv \mathcal{F}_{r0}^{(2)}$.

The usual thermodynamic integrals are defined in local equilibrium such that, 
\begin{eqnarray}
I_{nq} &\equiv &\frac{\left( -1\right) ^{q}}{\left( 2q+1\right) !!}
\int dK\,E_{\mathbf{k}}^{n-2q}\left( \Delta ^{\alpha \beta }k_{\alpha }k_{\beta}\right)^{q}f_{0\mathbf{k}}\;,  \label{I_nq} \\
J_{nq} &\equiv &\frac{\left( -1\right) ^{q}}{\left( 2q+1\right) !!}
\int dK\,E_{\mathbf{k}}^{n-2q}\left( \Delta ^{\alpha \beta }k_{\alpha }k_{\beta}\right)^{q}f_{0\mathbf{k}}
\left( 1-af_{0\mathbf{k}}\right) \;. \label{J_nq}
\end{eqnarray}
Here we also\ recall the following coefficients appearing in Eqs.\ (\ref{D_rho}) -- (\ref{D_rho_munu}), 
\begin{align}
\alpha _{r}^{\left( 0\right) }& \equiv \left( 1-r\right) I_{r1}-I_{r0}-\frac{n_{f0}}{D_{20}}\left( h_{0}G_{2r}-G_{3r}\right) \;,  
\label{alpha_i_0} \\
\alpha _{r}^{\left( 1\right) }& \equiv J_{r+1,1}-h_{0}^{-1}J_{r+2,1}\;,
\label{alpha_i_1} \\
\alpha _{r}^{\left( 2\right) }& \equiv I_{r+2,1}+\left( r-1\right) I_{r+2,2}\;,
\label{alpha_i_2} \\
\alpha _{r}^{h}& \equiv -\frac{\beta _{0}}{\varepsilon _{0}+P_{0}}J_{r+2,1}\;,
\label{alpha_i_h}
\end{align}
and 
\begin{eqnarray}
D_{nq} &\equiv &J_{n+1,q}J_{n-1,q}-J_{nq}^{2}\;, \\
G_{nm} &\equiv &J_{n,0}J_{m,0}-J_{n-1,0}J_{m+1,0}\;.
\end{eqnarray}

In the limit of a massless Boltzmann gas with constant cross section, 
$J_{nq}\equiv I_{nq}=\frac{\left( n+1\right) !}{2\left( 2q+1\right) !!}\beta_{0}^{2-n}P_{0}$, 
and thus $\alpha _{0}^{h} = -h_{0}^{-1}=-\beta _{0}/4$ 
as well $\alpha _{1}^{h} = 1$, hence the coefficients of interest are
\begin{eqnarray}
\gamma _{1}^{V(1)} &\equiv &\mathcal{F}_{10}^{\left( 1\right) }=\frac{2\beta_{0}}{3}\;,\; \; 
\gamma _{2}^{V(1)}\equiv \mathcal{F}_{20}^{\left( 1\right)}=\frac{\beta _{0}^{2}}{2}\;, \\
\gamma _{1}^{W(1)} &\equiv &\mathcal{F}_{11}^{\left( 1\right) }=-\frac{\beta_{0}^{2}}{12},\; \; 
\gamma _{2}^{W(1)}\equiv \mathcal{F}_{21}^{\left(1\right) }=-\frac{\beta _{0}^{3}}{12}\;, \\
\gamma _{1}^{(2)} &\equiv &\mathcal{F}_{10}^{\left( 2\right) }=\frac{\beta_{0}}{5}\;,\; \; 
\gamma _{2}^{(2)}\equiv \mathcal{F}_{20}^{\left( 2\right) }=\frac{\beta _{0}^{2}}{20}\;.
\end{eqnarray}

Also note that in the derivation of the relaxation equations we have expressed the proper-time
derivative and spatial derivative of the coefficients $\gamma^{(\ell)}_r$ by the following formulas
\begin{align}
\dot{\gamma}^{(\ell)}_r & = 
\frac{n_{0}}{D_{20}}\left[ \left( J_{20}\frac{\partial \gamma^{(\ell)}_r}{\partial \alpha _{0}} 
+ J_{10}\frac{\partial \gamma^{(\ell)}_r }{\partial \beta _{0}}\right) h_{0}
 -\left( J_{30}\frac{\partial \gamma^{(\ell)}_r }{\partial \alpha _{0}}+J_{20}
\frac{\partial \gamma^{(\ell)}_r}{\partial \beta _{0}}\right) \right] \theta \; , \\
\nabla ^{\mu } \gamma^{(\ell)}_r  & = 
\left( \frac{\partial \gamma^{(\ell)}_r}{\partial \alpha _{0}} + h_{0}^{-1}
\frac{\partial \gamma^{(\ell)}_r }{\partial \beta _{0}}\right) \nabla ^{\mu}\alpha _{0}  
- \beta_{0} \frac{\partial \gamma^{(\ell)}_r }{\partial \beta _{0}} \dot{u}^{\mu } 
+ \frac{\beta_{0}}{h_0} \frac{\partial \gamma^{(\ell)}_r }{\partial \beta _{0}} 
\textswab{q} E^{\mu } \; .
\end{align}
These equations follow from Eqs.\ (\ref{D_alpha}) -- (\ref{D_u_mu}) neglecting terms proportional to the 
dissipative fields and/or their derivatives.

\section{Transport coefficients}
\label{AppB}

\subsection{Coefficients of the bulk equation}

Noting that $\tau _{00}^{\left( 0\right) } = 1/\mathcal{A}_{00}^{\left(0\right) }$,
the transport coefficients found in the relaxation equation for the bulk
viscosity, Eq.\ (\ref{relax_bulk}), up to terms $N_{2}=0,\neq 1, 2$, are
\begin{equation}
\delta_{\Pi \Pi }=\frac{\tau _{00}^{\left( 0\right) }}{3} \left(2
+m_{0}^{2}\gamma _{2}^{\left( 0\right) } 
- m_{0}^{2}\frac{G_{20}}{D_{20}}\right) \;, \; \;
\lambda_{\Pi \pi }=\frac{m_{0}^{2}}{3}\tau_{00}^{\left( 0\right) }
\left( \gamma _{2}^{\left( 2\right) }-\frac{G_{20}}{D_{20}}\right) \;.
\end{equation}
Furthermore, the coefficients proportional to $V^{\mu}$ are
\begin{eqnarray}
\ell_{\Pi V} &=&-\frac{m_{0}^{2}}{3}\tau _{00}^{\left( 0\right) }\left(
\gamma _{1}^{V(1)}-\frac{G_{30}}{D_{20}}\right)\;  ,\; \; 
\tau_{\Pi V}=\frac{m_{0}^{2}}{3}\tau _{00}^{\left( 0\right) }\left( \beta _{0}\frac{\partial
\gamma _{1}^{V(1)}}{\partial \beta _{0}}-\frac{G_{30}}{D_{20}}\right)\;  , \\
\lambda_{\Pi V} &=&-\frac{m_{0}^{2}}{3}\tau _{00}^{\left( 0\right) }\left( 
\frac{\partial \gamma _{1}^{V(1)}}{\partial \alpha _{0}}+h_{0}^{-1}\frac{%
\partial \gamma _{1}^{V(1)}}{\partial \beta _{0}}\right)\;  ,
\end{eqnarray}%
while the new coefficients proportional to $W^{\mu }$ are
\begin{eqnarray}
\ell_{\Pi W} &=&-\frac{m_{0}^{2}}{3}\tau _{00}^{\left( 0\right) }\left(
\gamma _{1}^{W(1)}+\frac{G_{20}}{D_{20}}\right) ,\; \;  
\tau_{\Pi W}=\frac{m_{0}^{2}}{3}\tau _{00}^{\left( 0\right) }
\left( \beta _{0}\frac{\partial\gamma _{1}^{W(1)}}{\partial \beta _{0}}
+2\frac{G_{20}}{D_{20}}\right) , \\
\lambda_{\Pi W} &=&-\frac{m_{0}^{2}}{3}\tau _{00}^{\left( 0\right) }\left( 
\frac{\partial \gamma _{1}^{W(1)}}{\partial \alpha _{0}}
+h_{0}^{-1}\frac{\partial \gamma _{1}^{W(1)}}{\partial \beta _{0}}\right) .
\end{eqnarray}
The coefficients of the electric field are
\begin{equation}
\delta_{\Pi VE}=\frac{m_{0}^{2}}{3}\tau _{00}^{\left( 0\right) }
\left(\gamma _{2}^{V\left( 1\right) }-\frac{G_{20}}{D_{20}} 
-\frac{\beta_0}{h_0} \frac{\partial \gamma_{1}^{V(1)} }{\partial \beta_0} \right) \; ,\; \; 
\delta_{\Pi WE}=\frac{m_{0}^{2}}{3}\tau _{00}^{\left( 0\right) }
\left(\gamma_{2}^{W\left( 1\right) } 
-\frac{\beta_0}{h_0} \frac{\partial \gamma_{1}^{W(1)} }{\partial \beta_0}  \right)\; .
\end{equation}

\subsection{Coefficients of the diffusion equation}

Similarly with $\tau _{00}^{\left( 1\right) } = 1/\mathcal{A}_{00}^{\left(1\right) }$, 
the transport coefficients found in Eq.\ (\ref{relax_heat}) are
\begin{equation}
\ell_{V\Pi } = \tau _{00}^{\left( 1\right) }\left( h^{-1}_{0} 
- \gamma _{1}^{(0)}\right)\;  ,\; \; 
\tau _{V\Pi } = \tau_{00}^{\left( 1\right) }\left( h^{-1}_{0} -\beta _{0}\frac{\partial
\gamma_{1}^{(0)}}{\partial \beta _{0}}\right)\;  ,\; \;  
\lambda_{V\Pi }=\tau _{00}^{\left( 1\right) }\left( \frac{\partial \gamma _{1}^{(0)}}
{\partial \alpha _{0}}+h_{0}^{-1}\frac{\partial \gamma _{1}^{(0)}}{\partial\beta _{0}}\right)\;  .
\end{equation}
The coefficients proportional to $\pi ^{\mu \nu }$ are
\begin{equation}
\ell_{V\pi }=\tau _{00}^{\left( 1\right) }\left( h^{-1}_{0} - \gamma _{1}^{(2)}\right) \; ,\; \; 
\tau_{V\pi }=\tau _{00}^{\left( 1\right) }\left(
h^{-1}_{0} -\beta _{0}\frac{\partial \gamma _{1}^{(2)}}{\partial \beta_{0}}\right)\;  ,\; \;
\lambda_{V\pi }=\tau _{00}^{\left( 1\right)}
\left( \frac{\partial \gamma _{1}^{(2)}}{\partial \alpha _{0}}+h_{0}^{-1}
\frac{\partial \gamma _{1}^{(2)}}{\partial \beta _{0}}\right) \; .
\end{equation}%
The coefficients proportional to $V^{\mu}_f$ are
\begin{equation}
\delta_{VV} = \tau _{00}^{\left( 1\right) }\left(1+\frac{m_{0}^{2}}{3}
\gamma_{2}^{V\left( 1\right) }\right) ,\; \;  
\lambda _{VV}\equiv \frac{1}{5}\tau _{00}^{\left( 1\right) }
\left( 3+2m_{0}^{2}\gamma _{2}^{V\left(1\right) }\right) ,
\end{equation}%
while similarly the coefficients proportional to $W^{\mu }$ are
\begin{equation}
\delta _{WW}= \frac{\tau _{00}^{\left( 1\right) }}{3}\left( -4h^{-1}_{0}
+m_{0}^{2} \gamma _{2}^{W\left( 1\right) }\right) \; ,\; \;
\lambda _{WW}= \tau _{00}^{\left( 1\right) }\left( -h^{-1}_{0}+\frac{2m_{0}^{2}}{5}
\gamma _{2}^{W\left( 1\right) }\right) \; .
\end{equation}

The coefficients due to the magnetic field are
\begin{equation}
\delta _{VB}=\tau _{00}^{\left( 1\right) }\left( -h^{-1}_{0} + \gamma_{1}^{V(1)}\right)\; ,\; \; 
\delta _{WB}=\tau _{00}^{\left( 1\right) }\gamma_{1}^{W(1)}/;,
\end{equation}%
while the new coefficients due to the electric field are
\begin{eqnarray}
\delta _{VE} &=& \tau _{00}^{\left( 1\right) }\left( -n_{0} h^{-1}_{0}
+\beta_{0}J_{11}\right)\; , \\
\delta _{V\Pi E}&=&-\frac{1}{m_{0}^{2}}\tau_{00}^{\left( 1\right) }\left( 2+m_{0}^{2}\gamma _{2}^{(0)} - 
m_0^2\frac{\beta_0}{h_0} \frac{\partial \gamma_{1}^{\left( 0\right)}}{\partial \beta_0} \right)\; ,\ \ \
\delta _{V\pi E} =\tau _{00}^{\left( 1\right) } \left(
\gamma _{2}^{\left( 2\right) } - \frac{\beta_0}{h_0} 
\frac{\partial \gamma_{1}^{\left( 2\right)}}{\partial \beta_0} \right)\;.
\end{eqnarray}

\subsection{Coefficients of the shear-stress equation}

Finally, using $\tau _{00}^{\left( 2\right) } = 1/\mathcal{A}_{00}^{\left(2\right) }$, 
the transport coefficients found in Eq.\ (\ref{relax_shear}) are
\begin{equation}
\delta_{\pi \pi }=\tau _{00}^{\left( 2\right) }\left( \frac{4}{3}+\gamma_{2}^{\left( 2\right) }\frac{m_{0}^{2}}{3}\right) \;,\; \;  
\tau _{\pi \pi}=\frac{2\tau _{00}^{\left( 2\right) }}{7}\left( 5+2 m_{0}^{2}\gamma _{2}^{\left( 2\right) }\right)\; ,\; \; 
\lambda _{\pi \Pi }=\frac{2}{5}\tau _{00}^{\left( 2\right) }\left( 3+m_{0}^{2}\gamma_{2}^{\left( 0\right) }\right)\; .
\end{equation}%
Furthermore we have the coefficients of $V^{\mu}_f$,
\begin{equation}
\ell_{\pi V} = -\frac{2m_{0}^{2}}{5}\tau _{00}^{\left( 2\right) }
\gamma_{1}^{V\left( 1\right) }\;, \; \; 
\lambda_{\pi V}=-\frac{2m_{0}^{2}}{5}\tau_{00}^{\left( 2\right) }
\left( \frac{\partial \gamma _{1}^{V\left( 1\right) }}{\partial \alpha _{0}}
+h_{0}^{-1}\frac{\partial \gamma _{1}^{V\left(1\right) }}{\partial \beta _{0}}\right)\; ,\; \;  
\tau _{\pi V}= -\frac{2m_{0}^{2}}{5}\tau _{00}^{\left( 2\right) }
\beta _{0}\frac{\partial \gamma_{1}^{V\left( 1\right) }}{\partial \beta _{0}}\;,
\end{equation}%
and the new coefficients of $W^{\mu }$,
\begin{equation}
\ell_{\pi W} = \frac{2}{5}\tau _{00}^{\left( 2\right) }
\left(1 - m_{0}^{2}\gamma _{1}^{W\left( 1\right) } \right)\;,\; \; 
\lambda_{\pi W} =-\frac{2m_{0}^{2}}{5}\tau _{00}^{\left( 2\right)}
\left( \frac{\partial \gamma _{1}^{W\left( 1\right)}}{\partial \alpha _{0}}+h_{0}^{-1}
\frac{\partial \gamma _{1}^{W\left( 1\right) }}{\partial \beta _{0}}\right)\;, \; \;
\tau_{\pi W} = 2\tau _{00}^{\left( 2\right) } \left( 1-\frac{m^2_0}{5}\beta _{0}
\frac{\partial \gamma _{1}^{W\left(1\right) }}{\partial \beta _{0}} \right)\;.
\end{equation}

The new coefficient due to the magnetic and electric fields are
\begin{equation}
\delta _{\pi B} = 2\tau _{00}^{\left( 2\right) }\gamma _{1}^{\left( 2\right) }\;,
\end{equation}%
and 
\begin{equation}
\delta_{\pi VE} = \frac{2\tau _{00}^{\left( 2\right) }}{5}\left( 4+ m_{0}^{2}
\gamma_{2}^{V\left( 1\right) } -m_{0}^{2} \frac{\beta_0}{h_0} 
\frac{\partial \gamma_{1}^{V\left( 1\right)}}{\partial \beta_0} \right)\;, \; \;
\delta_{\pi WE}=\frac{2 m_0^2 \tau_{00}^{\left( 2\right) } }{5}
\left( \gamma_{2}^{W\left( 1\right) } 
-  \frac{\beta_0}{h_0} 
\frac{\partial\gamma_{1}^{W\left( 1\right)}}{\partial \beta_0} \right)\;.
\end{equation}

\end{document}